\documentclass{article}
\usepackage{spconf,amsmath,graphicx}
\usepackage{multirow}
\usepackage{subfigure}
\usepackage{graphics}
\usepackage{makecell}
\usepackage{verbatim}
\usepackage{url}
\usepackage{booktabs}
\usepackage{soul}
\usepackage{xcolor}

\title{Improving Speech Prosody of Audiobook Text-to-Speech Synthesis with Acoustic and Textual Contexts}
\name{\begin{tabular}{c} 
    Detai Xin$^{1,2}$\sthanks{This work was supported by JST SPRING, Grant Number JPMJSP2108.}, Sharath Adavanne$^{1}$, Federico Ang$^{1}$, Ashish Kulkarni$^{1}$, \\
     Shinnosuke Takamichi$^{2}$, Hiroshi Saruwatari$^{2}$\\
\end{tabular}
\vspace{-5mm}
}
\address{Rakuten Group, Inc., Japan$^{1}$,\\ Graduate School of Information Science and Technology, The University of Tokyo, Japan$^{2}$}
\begin{document}
\ninept
\maketitle
\allowdisplaybreaks
\begin{abstract}
We present a multi-speaker Japanese audiobook text-to-speech (TTS) system that leverages multimodal context information of preceding acoustic context and bilateral textual context to improve the prosody of synthetic speech.
Previous work either uses unilateral or single-modality context, which does not fully represent the context information.
The proposed method uses an acoustic context encoder and a textual context encoder to aggregate context information and feeds it to the TTS model, which enables the model to predict context-dependent prosody.
We conducted comprehensive objective and subjective evaluations on a multi-speaker Japanese audiobook dataset.
Experimental results demonstrate that the proposed method significantly outperforms two previous works.
Additionally, we present insights about the different choices of context - modalities, lateral information and length - for audiobook TTS that have never been discussed in the literature before.
\end{abstract}
\vspace{-1mm}
\section{Introduction}
\vspace{-1mm}
Recent text-to-speech (TTS) systems based on deep neural networks (DNNs) have been able to synthesize natural read-out speech~\cite{shen2018natural,ren2019fastspeech, ren2020fastspeech2}.
However, how to synthesize speech with a lot of prosody variations like audiobooks remains unsolved.
Synthesizing such speech requires the system to not only transform linguistic information but also para-/non-linguistic information such as emotions, and intentions from text to speech~\cite{taylor2009text,schuller2013paralinguistics, cole2015prosody}.
Speech prosody in audiobooks produced by professional speakers depends on several factors including characteristics, context, and styles (narration or dialogue)~\cite{Simon2018}.
Among these factors, context, either acoustic or textual, is popularly utilized in the literature to improve the prosody of audiobook TTS~\cite{oplustil2020using, gallegos2021comparing, xu2021improving, nakata21audiobook}.
This is because (1) consecutive utterances always have sequential relations in audiobooks; (2) unlike characteristics and other factors, context requires no additional cost to get.

However, existing work either (1) uses only single-modality context or (2) uses unilateral context which cannot fully leverage the power of context information.
Gallegos et al. first proposed to use acoustic context to improve the prosody of audiobook TTS~\cite{oplustil2020using}.
In their following work, they further used acoustic and textual contexts in audiobook TTS~\cite{gallegos2021comparing}, but only preceding contexts were used.
Xu et al. first proposed to use pretrained bidirectional encoder representations from transformers (BERT)~\cite{devlin2018bert} to encode preceding and succeeding sentences to incorporate textual context information in audiobook TTS~\cite{xu2021improving}.
Nakata et al. also used BERT embeddings, but only in an implicit way by encoding the target sentences with the bilateral context~\cite{nakata21audiobook}.
All of these works only used textual context with one or two sentences without trying longer context.

In this paper, we present a multi-speaker Japanese audiobook TTS system that fully and explicitly utilizes preceding acoustic context and bilateral textual context to improve the prosody of synthetic speech.
The proposed method first uses an acoustic context encoder (ACE)~\cite{oplustil2020using} to encode preceding mel-spectrograms as acoustic context representations.
Moreover, we propose a textual context encoder (TCE) based on attention mechanisms to extract textual context representations from BERT embeddings of bilateral textual context.
These context representations are then fed to a multi-speaker TTS model to guide it to synthesize the target utterance with appropriate prosody.
We conducted comprehensive experiments with both objective and subjective evaluations to verify the effectiveness of the proposed method.
We further compared how different modalities, laterals, and lengths of context influence the prosody of audiobook TTS, which was never studied in previous work.
Our contributions are summarized as follows:
\begin{itemize} \itemsep -1mm 
    \item We propose a multi-speaker Japanese audiobook TTS system with acoustic and textual context encoding mechanisms.
    \item We conduct comprehensive objective and subjective experiments to show the effectiveness of the proposed method.
    \item We further conduct experiments to find the best combination of context modalities, laterals, and lengths for audiobook TTS. 
\end{itemize}
Audio samples are publicly available\footnote{\url{https://aria-k-alethia.github.io/2022rat-demo/}}.
\vspace{-3mm}
\section{Related Work}
\vspace{-1mm}
Methods of audiobook TTS can be roughly grouped into two categories: single-sentence and multi-sentence methods.
Single-sentence methods synthesize one utterance at a time, and improves speech prosody by feeding auxiliary features like context~\cite{oplustil2020using, gallegos2021comparing, xu2021improving}, emotions~\cite{pan2021chapter}, and features extracted by DNNs like variational autoencoder~\cite{nakata2022predicting, wu2022discourse} and global style tokens (GST)~\cite{stanton2018predicting}.
Sufficient and correct prosody information should be maintained to guarantee satisfactory speech quality in such methods.

Multi-sentence methods, on the other hand, synthesize multiple sentences at a time. The length of the target sequence ranges from three sentences~\cite{makarov2022simple} to a whole paragraph~\cite{xue2022paratts}.
While such methods are potentially better than single-sentence methods, it requires more memory and sophisticated training process.

The proposed method in this work is a single-sentence method utilizing both acoustic and textual contexts.

\vspace{-3mm}
\section{Proposed Method}
\vspace{-1mm}
\begin{figure}[t]
\begin{center}
\centerline{
\includegraphics[width=\columnwidth]{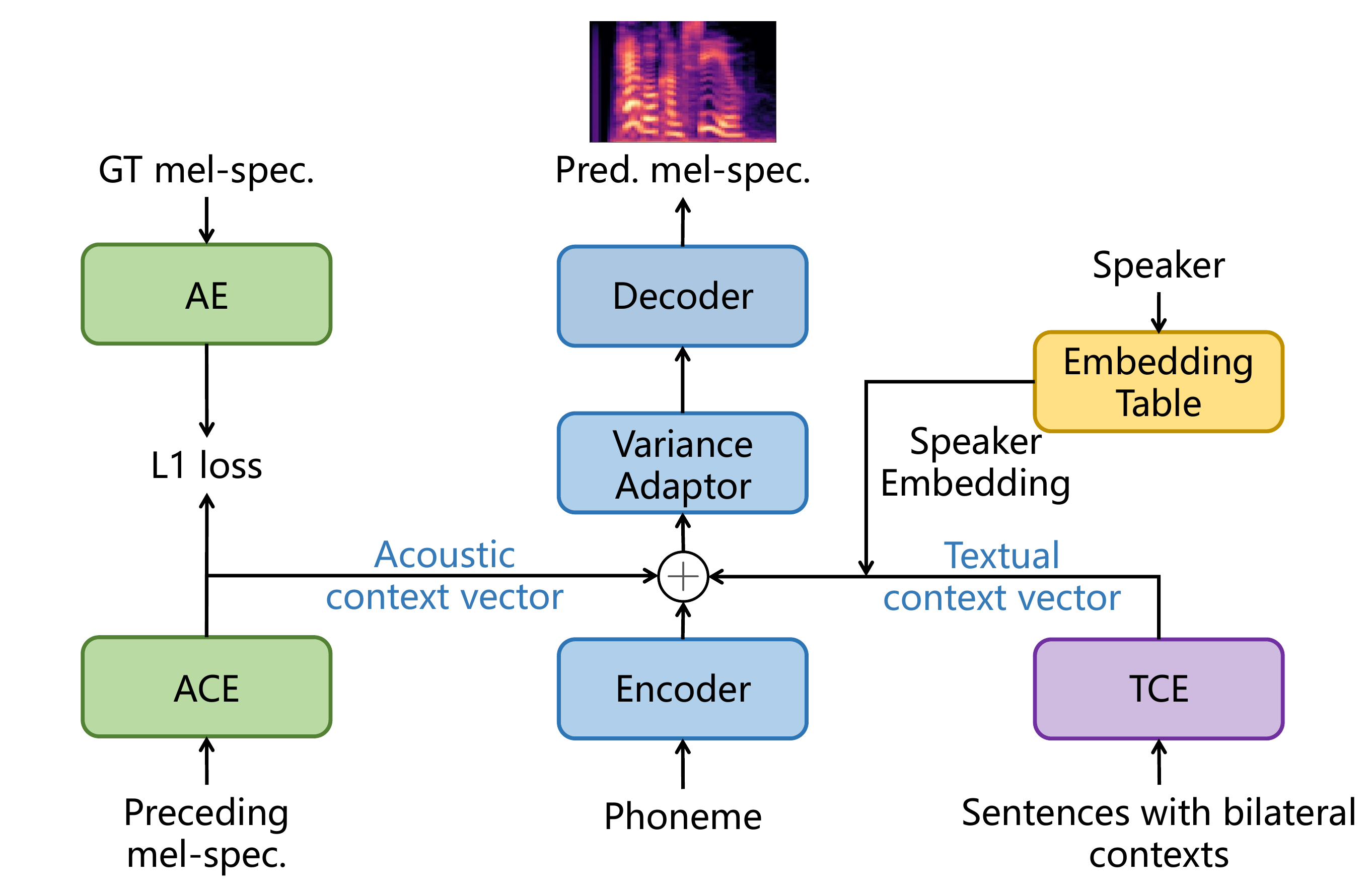}
}
\caption{Architecture of the proposed method.}
\label{figure:achitecture}
\vspace{-7mm}
\end{center}
\end{figure}
The general architecture of the proposed method is illustrated in Figure~\ref{figure:achitecture}.
The model contains three components: an improved multi-speaker FastSpeech2~\cite{ren2020fastspeech2} for mel-spectrograms synthesis, an acoustic context encoder and a textual context encoder for context encoding.
We introduce these components separately in the following sections.

\vspace{-3mm}
\subsection{Improved multi-speaker FastSpeech2}
\label{subsection:fs2}
We follow the original FastSpeech2~\cite{ren2020fastspeech2} to construct the TTS model that contains a phoneme encoder, a variance adaptor, and a mel-spectrograms decoder.
To adapt the model to a multi-speaker setting, we first create a look-up embedding table for speakers.
The speaker embedding is summed to the output of the phoneme encoder and fed to the following modules to generate mel-spectrograms for the corresponding speaker.
Second, we also use speaker-dependent pitch normalization~\cite{kharitonov2022text} to disentangle speaker information from pitch values.
Specifically, for a pitch value $p_{s}$ of speaker $s$ of a voiced frame, we normalize it to $\bar{p}_{s}$ by:
$
    \bar{p}_{s} = \frac{p_{s} - \mu_{s}}{\sigma_{s}},
$
where $\mu_{s}, \sigma_{s}$ are the mean and standard deviation values of the pitch of the speaker $s$, respectively.

We also notice that the utterances of audiobooks are longer than utterances in read-out corpus, but the absolute positional encoding~\cite{vaswani2017attention} used in the original FastSpeech2 cannot well handle long sequences.
To solve this problem, we replace the absolute positional encoding with relative positional encoding~\cite{shaw2018self} so that the model can handle sequences with any length.
We follow the previous work~\cite{shaw2018self} and add relative positional embeddings to the attention layers of both the encoder and the decoder.
The clipping distance is set to $4$ so that the model can capture relative position differences within this value.
For more details please refer to the original paper.

Finally, the original FastSpeech2 uses a length regulator in the variance adaptor before the pitch and energy predictors so that the model outputs frame-level pitch and energy, which makes the synthesized speech unstable.
Therefore we move the length regulator after the two predictors to learn phoneme-level pitch and energy.
In our preliminary experiments, we found all of the aforementioned modifications could improve the overall performance.

The deterministic nature of the above model makes it difficult to synthesize speech with various prosodies.
Therefore, the proposed method further uses two context encoders to incorporate context information into the model. 

\vspace{-3mm}
\subsection{Acoustic context encoding}
\begin{figure}[t]
\begin{center}
\centerline{
\includegraphics[width=\columnwidth]{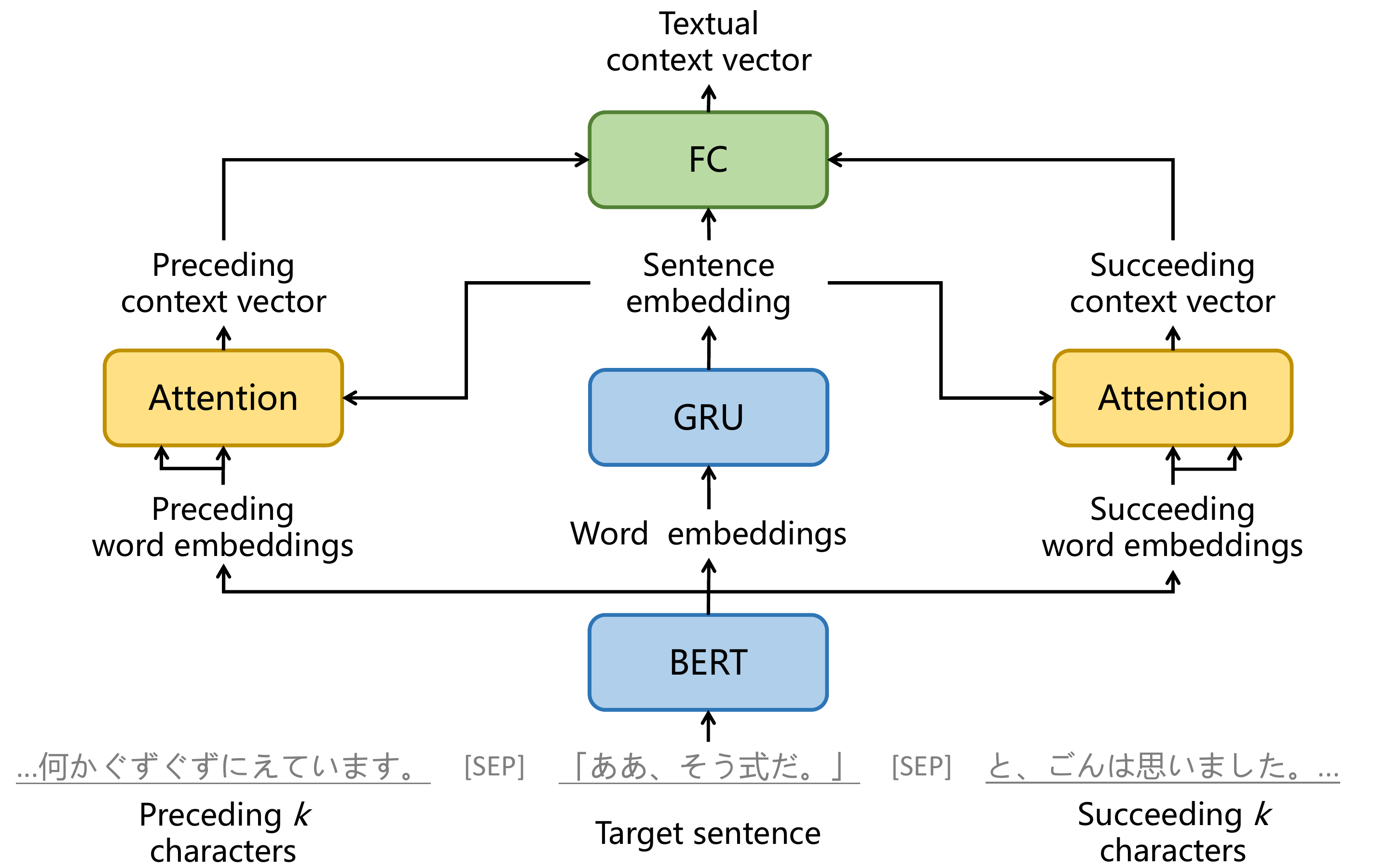}
}
\caption{Architecture of TCE. FC denotes fully connected layer.}
\label{figure:tce}
\vspace{-7mm}
\end{center}
\end{figure}
In audiobooks it can be assumed that the prosodies of consecutive utterances have minor differences, hence using acoustic context can intuitively make consecutive utterances more coherent in single-sentence audiobook TTS methods.
To this end, the proposed method uses ACE to encode acoustic context.
Since during inference, only preceding utterances are available, ACE only encodes preceding acoustic context.
Supposing the index of the target utterance for synthesis is $N$, we use GST as the implementation of ACE to extract a fixed-length acoustic context vector from the $(N-1)$-th mel-spectrogram.
The context vector is summed to the output of the phoneme encoder to assist the synthesis of the $N$-th mel-spectrogram.
Note that, although during training we use the ground truth (GT) $(N-1)$-th mel-spectrogram as the input of ACE, during inference we use the synthesized one as the input.

Following previous work~\cite{oplustil2020using}, we set an extra next-prediction task for ACE.
As Figure~\ref{figure:achitecture} illustrates, we use an acoustic encoder (AE) to extract a fixed-length vector from the $N$-th mel-spectrogram.
Here AE is implemented as another GST module that has different parameters from the one of ACE.
We then minimize the L1 distance between the two vectors extracted by ACE and AE.
While previous work doesn't explain why the extra task is effective~\cite{oplustil2020using}, we suppose this is because the extra task forces ACE to learn a corresponding relation between the $(N-1)$-th and the $N$-th mel-spectrograms.
During training, we add the L1 loss term to the loss function of the TTS model and jointly train the whole model.

\vspace{-3mm}
\subsection{Textual context encoding}
Textual context is also an informative source for producing appropriate prosody.
Phrases like ``a man/woman says", and ``happily/sadly" in the contexts can be useful cues to predict the prosody of the target sentence.
Therefore we propose a textual context encoder (TCE) to incorporate textual context information in the model.
The architecture of TCE is illustrated in Figure~\ref{figure:tce}.
First, the bilateral context and the target text are fed to a pretrained BERT model to extract word embeddings.
Note that, different from most previous works~\cite{gallegos2021comparing, xu2021improving, nakata21audiobook}, we set the context length of each lateral to $k$ characters instead of setting it to a certain number of sentences, by which we can easily evaluate model performance with different context lengths.
The word embeddings of the target sentence are then fed to a gated recurrent unit (GRU)~\cite{chung2014empirical}, and the last hidden state of GRU is used as the sentence embedding of the target sentence.
Next, the sentence embedding is used as a query vector in two attention modules with the word embeddings of preceding and succeeding contexts as keys and values to extract bilateral context vectors.
Finally, the two context vectors and the sentence embedding are concatenated together and fed to a fully connected (FC) layer to get the textual context vector.
This vector is then summed to the output of the phoneme encoder to incorporate textual context information in the TTS model.
\vspace{-5mm}
\section{Experiments}
\vspace{-5mm}
\subsection{Setup}
\vspace{-3mm}
We used J-MAC, a multi-speaker Japanese audiobook corpus produced by professional speakers, as the dataset~\cite{takamichi2022j}.
We selected $9$ speakers from J-MAC who at least have $3$ books to ensure each speaker has sufficient data for training.
We then randomly picked up gender-balanced $6$ speakers from the $9$ speakers as the test speakers and excluded one audiobook for each test speaker from the training set as the test set.
The final training set contained about 7 hours of audio data.
For the BERT model in TCE, we used "bert-base-japanese-v2"\footnote{\url{https://huggingface.co/cl-tohoku/bert-base-japanese-v2}} pretrained on $4$GB Japanese Wikipedia data.
We used the output of the last layer of the BERT model as the word embeddings used in TCE.
OpenJTalk\footnote{\url{https://open-jtalk.sp.nitech.ac.jp/}} was used to convert Japanese characters to phonemes.
For the forced alignment we used Julius~\cite{lee2001julius} to get the duration of each phoneme.
The pitch values were extracted by the WORLD vocoder~\cite{morise2016world}.
We used the pretrained ``\texttt{\detokenize{UNIVERSAL_V1}}" HiFi-GAN model\footnote{\url{https://github.com/jik876/hifi-gan}} to convert mel-spectrograms into time-domain waveforms.

For the TTS model we used the same parameter setting as the one of the previous work~\cite{ren2020fastspeech2} except for the modifications mentioned in Section~\ref{subsection:fs2}.
The dimension of the speaker embedding was set to $256$.
In ACE, the GST token number was set to $10$.
Following the original work~\cite{wang2018style}, we used multi-head attention with $8$ heads to improve the robustness.
In TCE, the number of hidden units in GRU was set to $256$.
The dimension of both the acoustic and the textual context vectors was set to $256$ so that they could be summed to the output of the phoneme encoder.

During training, the batch size was set to $32$.
We used Adam~\cite{kingma2014adam} as the optimizer, with a scheduled learning rate proposed in~\cite{vaswani2017attention}.
However, for the fine-tuning of the pre-trained BERT model in TCE, we set the learning rate to $10^{-7}$.
The combined model converged in around $200$k steps.

We trained several variations of the proposed model to study how different modalities, laterals, and lengths of context influence the prosody of the synthesized speech.
We denote the proposed method as ATCE-\{pre., suc., bi\}, where the suffix represents textual context laterals (pre. for preceding, suc. for succeeding, and bi for bilateral context).
We also trained the proposed models without ACE, which are denoted by TCE-\{pre., suc., bi\}.
We used two previous methods as the baselines.
The first baseline, denoted by ACE, uses ACE to utilize acoustic context~\cite{oplustil2020using}.
The second baseline uses one-sentence bilateral textual context implicitly by feeding the target sentence with one-sentence bilateral context to the BERT model but only inputting the word embeddings of the target sentence to the TTS model~\cite{nakata21audiobook}.
Since the original method of Nakata et al. didn't use FastSpeech2 to synthesize mel-spectrograms, we adapted it to the proposed method by (1) changing the context length to one sentence and (2) only inputting the target sentence embedding to the TTS model.

\vspace{-3mm}
\subsection{Objective evaluations}
\vspace{-1mm}
\subsubsection{Metrics}
\vspace{-1mm}
In the objective evaluations we used several metrics to evaluate the synthetic speech:
\begin{itemize} \itemsep -1mm 
    \item \textbf{Character error rate (CER)} computed using Vosk Japanese speech recognition API\footnote{\url{https://github.com/alphacep/vosk-api}}.
    \item \textbf{Mel-cepstral distortion (MCD)} computed with dynamic time warping (DTW).
    \item \textbf{F0 root mean square error (F0-RMSE)} computed with DTW.
    \item \textbf{Gross Pitch Error (GPE)} represents the proportion of voiced frames whose relative pitch error is higher than a certain threshold ($20\%$ in this work).
    \item \textbf{Accuracy of speaker classification (ACC)} computed by a speaker classifier trained on the training set.
\end{itemize}
Here CER and MCD measure the general speech quality, ACC measures the speaker similarity, F0-RMSE and GPE measure the performance on speech prosody.

\vspace{-3mm}
\subsubsection{Textual context length}
\vspace{-1mm}
\begin{table}[t]
    \centering
    \caption{Results of objective evaluation for the proposed ATCE-bi model with different context length $k$. \textbf{Bold} indicates the best score.}
    \vspace{1mm}
    \begin{tabular}{c|ccccc}
    \hline
    $k$ & CER($\downarrow$) & MCD($\downarrow$) & F0-RMSE($\downarrow$) & GPE($\downarrow$) & ACC($\uparrow$) \\
    \hline\hline
    $16$  & $0.206$ & $7.69$ & $28.60$ & $13.49$ & $99.18$ \\
    $32$ & $0.206$ & $7.68$ & $28.33$ & $13.21$ & $99.32$ \\
    $64$ & $0.208$ & $7.72$ & $\mathbf{28.19}$  & $\mathbf{12.90}$ & $98.77$ \\
    $128$ & $0.207$ & $7.64$ &  $28.56$ & $13.51$ & $98.77$ \\
    $128 \rightarrow 64$ & $0.206$ & $7.65$ &  $28.43$ & $13.23$ & $98.91$ \\
    \hline
    \end{tabular}
    \label{tab:objective_length}
    \vspace{-3mm}
\end{table}
We first evaluated model performances with different textual context length $k$.
We trained ATCE-bi with $k$ in $\{16, 32, 64, 128\}$.
The result is shown in Table~\ref{tab:objective_length}.
First, all models have similar CER, MCD, and ACC, which is natural since the proposed method only focuses on prosody.
Second, the best performance is obtained when $k=64$, which demonstrates that increasing textual context length can improve speech prosody, but when the length is too long ($k=128$), the performance degrades.
We suppose this is because textual context with a long distance to the target sentence is not relevant for predicting the target prosody.
To verify this hypothesis, we also set $k=64$ in the $k=128$ model for inference.
As expected, the performance increases slightly in this case ($128 \rightarrow 64$) in Table~\ref{tab:objective_length}), which implies the correctness of the hypothesis.
Given that the average sentence length of the corpus is $27$ characters, our results suggest that the best textual context length is about $2\text{-}3$ sentences from either lateral of the target sentence.

\vspace{-3mm}
\subsubsection{Context modalities and laterals}
\vspace{-1mm}
\begin{table}[t]
    \centering
    \caption{Results of objective evaluation for models with different types of context. \textbf{Bold} indicates the best score. $^{*}$ represents significant improvement over the baseline methods with p-value $< 0.05$.}
    \vspace{1mm}
    \begin{tabular}{c|ccccc}
    \hline
    Model & CER($\downarrow$) & MCD($\downarrow$) & F0-RMSE($\downarrow$) & GPE($\downarrow$) & ACC($\uparrow$) \\
    \hline \hline
    GT       & $0.192$ & N/A & N/A & N/A & $96.19$  \\
    HiFi-GAN & $0.208$ & $4.25$ & $14.03$ & $2.2$ & $98.64$ \\
    \hline
    ACE      & $0.207$ & $7.85$ & $29.05$ &  $14.16$  & $98.37$ \\
    Nakata et al.  & $0.206$ & $7.70$ & $28.89$ & $13.21$ & $98.64$ \\
    \hline
    TCE-pre. & $0.206$ & $7.71$ & $28.19$ & $13.00$ & $98.64$ \\
    TCE-suc. & $0.206$ & $7.70$ & $28.82$ & $13.18$ & $99.05$ \\
    TCE-bi & $0.207$ & $7.69$ & $28.84$ & $12.94$ & $98.50$ \\
    \hline
    ATCE-pre. & $0.205$ & $7.68$ & $28.86$ & $13.11$ & $99.18$ \\
    ATCE-suc. & $0.209$ & $\mathbf{7.66}$ & $\mathbf{27.93^{*}}$ & $\mathbf{12.57^{*}}$ & $98.91$\\
    ATCE-bi & $0.208$ & $7.72$ &  $28.19$ & $12.90$ & $98.77$ \\
    \hline
    \end{tabular}
    \label{tab:objective_type}
    \vspace{-3mm}
\end{table}
We then evaluated model performances with different context modalities and laterals.
All of the models were trained with $k=64$ obtained in the previous section.
The result is shown in Table~\ref{tab:objective_type}.
First, the proposed ATCE-* models outperform all the baseline models, which demonstrates the effectiveness of the proposed method.
Second, all TCE-* models outperform ACE, which demonstrates the effectiveness of introducing textual context.
Third, all ATCE-* models except ATCE-pre. have better performance than the corresponding TCE-* models, which demonstrates the necessity of combining both acoustic and textual context information for synthesis.
Finally, to our surprise, ATCE-suc. using succeeding textual context obtains the best performance and has better performance than ATCE-bi using bilateral textual context.
We suppose this is because the information overlapping between the $(N-1)$-th mel-spectrogram and the preceding texts makes succeeding textual context more informative for the model.
Therefore in such case using preceding acoustic and bilateral textual contexts together is probably not beneficial and can even confuse the TTS model.
This hypothesis can also be verified by comparing the performances of ATCE-pre. and TCE-pre..

We also notice that the accuracy of speaker classification of the GT model is the worst among all models.
We believe this is because the speakers usually change their voices to act different characters in the audiobooks, which makes it difficult to recognize the speaker identity.

\vspace{-3mm}
\subsubsection{Prosody under different contexts}
\vspace{-1mm}
\begin{figure}[t]
\begin{center}
\centerline{
\includegraphics[width=0.95\columnwidth]{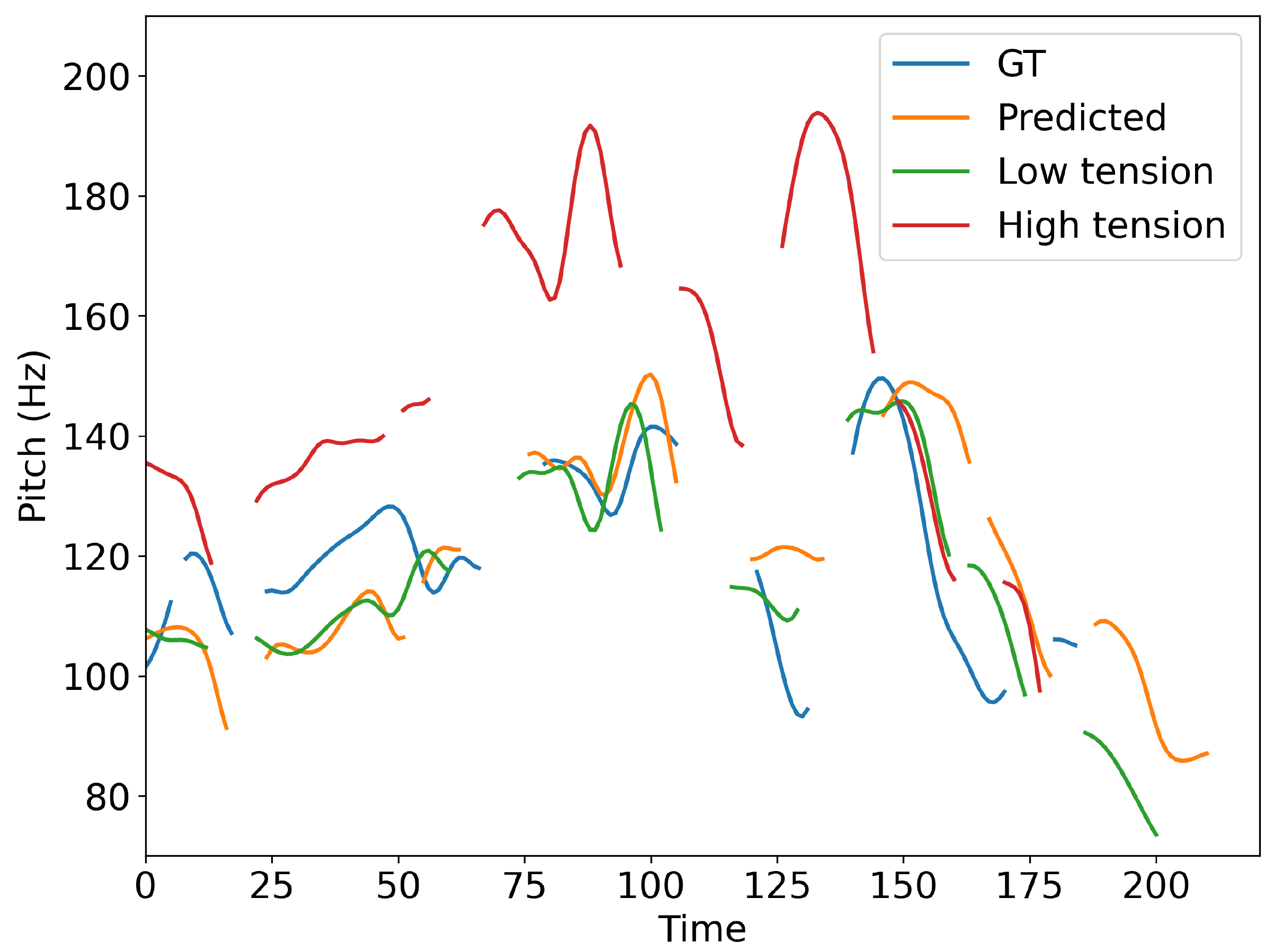}
}
\caption{Pitch contours of the same utterance synthesized using the ATCE-bi model with $2$ random contexts. ``Predicted'' represents the one synthesized with the correct context.}
\label{figure:prosody}
\vspace{-7mm}
\end{center}
\end{figure}
Finally, we verified whether the proposed method actually learned to predict context-dependent prosody.
We selected an utterance from the test set and synthesized it using the ATCE-bi model with random contexts.
We observed that the model could synthesize the same text with different tensions.
We selected typical examples and visualized their pitch contours in Figure~\ref{figure:prosody}.
It can be seen that the prosody varies a lot with different contexts, which proves that the proposed method can predict context-dependent speech prosody.

\vspace{-3mm}
\subsection{Subjective evaluations}
\vspace{-1mm}
\begin{table}[t]
    \centering
    \caption{Results of MOS evaluation. \textbf{Bold} indicates the best method without overlapping $95$\% confidence interval.}
    \vspace{1mm}
    \begin{tabular}{c|c}
    \hline
    Model              & MOS  \\
    \hline
    ACE                & $3.26$ \\
    Nakata et al.      & $3.30$ \\
    \hline
    ATCE-suc.          & $\mathbf{3.38}$ \\
    ATCE-bi           & $3.35$           \\
    \hline
    \end{tabular}
    \label{tab:mos_type}
    \vspace{-3mm}
\end{table}
\subsubsection{Multi-sentence speech naturalness MOS test}
In the subjective evaluations, we first conducted a standard $5$-scale mean opinion score (MOS) test.
We fine-tuned the HiFi-GAN vocoder on the training set for $3000$ epochs with an initial learning rate $10^{-5}$.
Following previous work~\cite{takamichi2022j}, we conducted a five-sentence MOS test, in which the listeners rate the naturalness of an audio including five consecutive utterances.
In this test we selected the two baselines and ATCE-\{suc., bi.\} with the best objective performance obtained in the previous section to evaluate.
For each test speaker we synthesized $10$ five-sentence audios, in which we inserted a $0.5$ second pause after each sentence.
The duration of each audio ranges from $40$ seconds to $1$ minute.
We used Lancers\footnote{\url{https://www.lancers.jp/}}, a Japanese crowd-sourcing platform, to conduct the test.
$32$ listeners joined in the test.
Each listener rated $30$ audios with $5$ dummy samples at the beginning whose ratings were not counted in the final result.
Each audio had $3$ answers on average.

The result is shown in Table~\ref{tab:mos_type}.
It can be seen that all proposed models outperform the two baseline models, and ATCE-suc. obtains the best performance, which is consistent with the results of the objective evaluations.
This again demonstrates the effectiveness of the proposed method.

\vspace{-3mm}
\subsubsection{Preference AB test}
\vspace{-1mm}
Next we conducted a preference AB test using the same five-sentence audios synthesized in the previous section.
We selected two AB pairs: (ATCE-suc., ACE), (ATCE-suc, Nakata et al.).
$40$ listeners participated in the test.
Each listener rated $10$ pairs, in which $5$ pairs have the same audios but different orders from the rest $5$ audios.
Each pair had $3$ answers on average.
\begin{table}[t]
    \centering
    \caption{Results of AB preference evaluation. \textbf{Bold} indicates the best method with p-value $< 0.05$.}
    \vspace{1mm}
    \begin{tabular}{c|cc|c}
    \hline
    Method A & Score A & Score B & Method B  \\
    \hline
    ATCE-suc. & $\mathbf{0.615}$ & $0.385$ & ACE \\
    ATCE-suc. & $\mathbf{0.545}$ & $0.455$ & Nakata et al. \\
    \hline
    \end{tabular}
    \label{tab:ab}
    \vspace{-3mm}
\end{table}
The result is shown in Table~\ref{tab:ab}.
It can be seen that the proposed ATCE-suc. method significantly outperforms the two baselines, which is consistent with the results obtained in the previous section.
All in all, the proposed method utilizing informative acoustic and textual contexts obtains the best performance in all evaluations.
\vspace{-1mm}
\section{Conclusions}
\vspace{-1mm}
This paper presented a Japanese multi-speaker audiobook TTS system that fully and explicitly utilized preceding acoustic context and bilateral textual context to improve the prosody of the synthetic speech.
Experimental results demonstrated that the proposed method significantly outperformed two previous work in both objective and subjective evaluations.
We also found it was helpful to use multimodal contexts and the optimal textual length was about $2\text{-}3$ sentences.
These results can potentially shed light on future researches in this field.
Future work could be extending ACE and TCE to frame-level resolution.

\bibliographystyle{IEEEbib}
\bibliography{refs}

\begin{thebibliography}{10}

\bibitem{shen2018natural}
Jonathan Shen, Ruoming Pang, Ron~J Weiss, Mike Schuster, Navdeep Jaitly,
  Zongheng Yang, Zhifeng Chen, Yu~Zhang, Yuxuan Wang, Rj~Skerrv-Ryan, et~al.,
\newblock ``Natural tts synthesis by conditioning wavenet on mel spectrogram
  predictions,''
\newblock in {\em Proc. ICASSP}. IEEE, 2018, pp. 4779--4783.

\bibitem{ren2019fastspeech}
Yi~Ren, Yangjun Ruan, Xu~Tan, Tao Qin, Sheng Zhao, Zhou Zhao, and Tie-Yan Liu,
\newblock ``Fastspeech: Fast, robust and controllable text to speech,''
\newblock {\em Proc. NeurIPS}, vol. 32, 2019.

\bibitem{ren2020fastspeech2}
Yi~Ren, Chenxu Hu, Xu~Tan, Tao Qin, Sheng Zhao, Zhou Zhao, and Tie-Yan Liu,
\newblock ``Fastspeech 2: Fast and high-quality end-to-end text to speech,''
\newblock {\em arXiv preprint arXiv:2006.04558}, 2020.

\bibitem{taylor2009text}
Paul Taylor,
\newblock {\em Text-to-speech synthesis},
\newblock Cambridge university press, 2009.

\bibitem{schuller2013paralinguistics}
Bj{\"o}rn Schuller, Stefan Steidl, Anton Batliner, Felix Burkhardt, Laurence
  Devillers, Christian M{\"u}Ller, and Shrikanth Narayanan,
\newblock ``Paralinguistics in speech and language—state-of-the-art and the
  challenge,''
\newblock {\em Computer Speech \& Language}, vol. 27, no. 1, pp. 4--39, 2013.

\bibitem{cole2015prosody}
Jennifer Cole,
\newblock ``Prosody in context: A review,''
\newblock {\em Language, Cognition and Neuroscience}, vol. 30, no. 1-2, pp.
  1--31, 2015.

\bibitem{Simon2018}
King Simon, Crumlish Jane, Martin Amy, and Wihlborg Lovisa,
\newblock ``The blizzard challenge 2018,''
\newblock in {\em Proc. Blizzard Challenge workshop}, 2018.

\bibitem{oplustil2020using}
Pilar Oplustil-Gallegos and Simon King,
\newblock ``Using previous acoustic context to improve text-to-speech
  synthesis,''
\newblock {\em arXiv preprint arXiv:2012.03763}, 2020.

\bibitem{gallegos2021comparing}
Pilar~Oplustil Gallegos, Johannah O'Mahony, and Simon King,
\newblock ``Comparing acoustic and textual representations of previous
  linguistic context for improving text-to-speech,''
\newblock in {\em The 11th ISCA Speech Synthesis Workshop (SSW11)}. ISCA, 2021,
  pp. 205--210.

\bibitem{xu2021improving}
Guanghui Xu, Wei Song, Zhengchen Zhang, Chao Zhang, Xiaodong He, and Bowen
  Zhou,
\newblock ``Improving prosody modelling with cross-utterance bert embeddings
  for end-to-end speech synthesis,''
\newblock in {\em Proc. ICASSP}. IEEE, 2021, pp. 6079--6083.

\bibitem{nakata21audiobook}
Wataru Nakata, Tomoki Koriyama, Shinnosuke Takamichi, Naoko Tanji, Yusuke
  Ijima, Ryo Masumura, and Hiroshi Saruwatari,
\newblock ``Audiobook speech synthesis conditioned by cross-sentence
  context-aware word embeddings,''
\newblock in {\em Proc. 11th ISCA Speech Synthesis Workshop (SSW 11)}, 2021,
  pp. 211--215.

\bibitem{devlin2018bert}
Jacob Devlin, Ming-Wei Chang, Kenton Lee, and Kristina Toutanova,
\newblock ``Bert: Pre-training of deep bidirectional transformers for language
  understanding,''
\newblock {\em arXiv preprint arXiv:1810.04805}, 2018.

\bibitem{pan2021chapter}
Junjie Pan, Lin Wu, Xiang Yin, Pengfei Wu, Chenchang Xu, and Zejun Ma,
\newblock ``A chapter-wise understanding system for text-to-speech in chinese
  novels,''
\newblock in {\em Proc. ICASSP}. IEEE, 2021, pp. 6069--6073.

\bibitem{nakata2022predicting}
Wataru Nakata, Tomoki Koriyama, Shinnosuke Takamichi, Yuki Saito, Yusuke Ijima,
  Ryo Masumura, and Hiroshi Saruwatari,
\newblock ``{Predicting VQVAE-based Character Acting Style from
  Quotation-Annotated Text for Audiobook Speech Synthesis},''
\newblock {\em Proc. Interspeech}, pp. 4551--4555, 2022.

\bibitem{wu2022discourse}
Ning-Qian Wu, Zhao-Ci Liu, and Zhen-Hua Ling,
\newblock ``Discourse-level prosody modeling with a variational autoencoder for
  non-autoregressive expressive speech synthesis,''
\newblock in {\em Proc. ICASSP}. IEEE, 2022, pp. 7592--7596.

\bibitem{stanton2018predicting}
Daisy Stanton, Yuxuan Wang, and RJ~Skerry-Ryan,
\newblock ``Predicting expressive speaking style from text in end-to-end speech
  synthesis,''
\newblock in {\em 2018 IEEE Spoken Language Technology Workshop (SLT)}. IEEE,
  2018, pp. 595--602.

\bibitem{makarov2022simple}
Peter Makarov, Ammar Abbas, Mateusz {\L}ajszczak, Arnaud Joly, Sri Karlapati,
  Alexis Moinet, Thomas Drugman, and Penny Karanasou,
\newblock ``Simple and effective multi-sentence tts with expressive and
  coherent prosody,''
\newblock {\em arXiv preprint arXiv:2206.14643}, 2022.

\bibitem{xue2022paratts}
Liumeng Xue, Frank~K Soong, Shaofei Zhang, and Lei Xie,
\newblock ``Paratts: Learning linguistic and prosodic cross-sentence
  information in paragraph-based tts,''
\newblock {\em IEEE/ACM Transactions on Audio, Speech, and Language
  Processing}, vol. 30, pp. 2854--2864, 2022.

\bibitem{kharitonov2022text}
Eugene Kharitonov, Ann Lee, Adam Polyak, Yossi Adi, Jade Copet, Kushal
  Lakhotia, Tu~Anh Nguyen, Morgane Riviere, Abdelrahman Mohamed, Emmanuel
  Dupoux, et~al.,
\newblock ``Text-free prosody-aware generative spoken language modeling,''
\newblock in {\em Proceedings of the 60th Annual Meeting of the Association for
  Computational Linguistics (Volume 1: Long Papers)}, 2022, pp. 8666--8681.

\bibitem{vaswani2017attention}
Ashish Vaswani, Noam Shazeer, Niki Parmar, Jakob Uszkoreit, Llion Jones,
  Aidan~N Gomez, {\L}ukasz Kaiser, and Illia Polosukhin,
\newblock ``Attention is all you need,''
\newblock {\em Proc. NeurIPS}, vol. 30, 2017.

\bibitem{shaw2018self}
Peter Shaw, Jakob Uszkoreit, and Ashish Vaswani,
\newblock ``Self-attention with relative position representations,''
\newblock {\em arXiv preprint arXiv:1803.02155}, 2018.

\bibitem{chung2014empirical}
Junyoung Chung, Caglar Gulcehre, Kyunghyun Cho, and Yoshua Bengio,
\newblock ``Empirical evaluation of gated recurrent neural networks on sequence
  modeling,''
\newblock in {\em NIPS 2014 Workshop on Deep Learning}, 2014.

\bibitem{takamichi2022j}
Shinnosuke Takamichi, Wataru Nakata, Naoko Tanji, and Hiroshi Saruwatari,
\newblock ``J-mac: Japanese multi-speaker audiobook corpus for speech
  synthesis,''
\newblock {\em arXiv preprint arXiv:2201.10896}, 2022.

\bibitem{lee2001julius}
Akinobu Lee, Tatsuya Kawahara, and Kiyohiro Shikano,
\newblock ``Julius---an open source real-time large vocabulary recognition
  engine,''
\newblock 2001.

\bibitem{morise2016world}
Masanori Morise, Fumiya Yokomori, and Kenji Ozawa,
\newblock ``World: a vocoder-based high-quality speech synthesis system for
  real-time applications,''
\newblock {\em IEICE TRANSACTIONS on Information and Systems}, vol. 99, no. 7,
  pp. 1877--1884, 2016.

\bibitem{wang2018style}
Yuxuan Wang, Daisy Stanton, Yu~Zhang, RJ-Skerry Ryan, Eric Battenberg, Joel
  Shor, Ying Xiao, Ye~Jia, Fei Ren, and Rif~A Saurous,
\newblock ``Style tokens: Unsupervised style modeling, control and transfer in
  end-to-end speech synthesis,''
\newblock in {\em Proc. ICML}. PMLR, 2018, pp. 5180--5189.

\bibitem{kingma2014adam}
Diederik~P Kingma and Jimmy Ba,
\newblock ``Adam: A method for stochastic optimization,''
\newblock {\em arXiv preprint arXiv:1412.6980}, 2014.

\end{thebibliography}

\end{document}